\documentclass[preprint,12pt]{elsarticle}
\usepackage{graphicx}
\usepackage{amssymb}
\usepackage{amsmath}
\usepackage{hyperref}

\journal{Physica A}

\begin{document}

\begin{frontmatter}

\title{Two-Population Dynamics in a Growing Network Model }
  
\author{Kristinka Ivanova\corref{cor1}}
\ead{ivanova@psu.edu}
\cortext[cor1]{Corresponding author}
\address{The Pennsylvania State 
University, University Park, PA 16802, USA}

\author{Ivan Iordanov}
\ead{ioi100@psu.edu}
\address{Department of Physics, The Pennsylvania State 
University, University Park, PA 16802, USA}

\begin{abstract}
We introduce a growing network evolution 
model with nodal attributes. 
The model describes the interactions 
between potentially violent $V$ and non-violent $N$ agents who have 
different affinities in establishing connections within their own population 
versus between the populations. 
The model is able to generate all stable triads observed in real 
social systems.
In the framework of rate equations theory, we employ the mean-field approximation
to derive analytical expressions of the degree distribution and the local
clustering coefficient for each type of nodes.
Analytical derivations agree well with numerical simulation results.
The assortativity of the potentially
violent network qualitatively resembles the connectivity pattern in terrorist
networks that was recently reported.
The assortativity of the network driven by aggression shows clearly 
different behavior than the assortativity of the networks with connections 
of non-aggressive nature in agreement with recent empirical results of an
online social system.

\end{abstract}


\end{frontmatter}

\section{Introduction}

Complex networks have been the focus of the study of dynamical properties 
of complex systems in nature and society in the last decade  
\cite{Watts98,AB2002,newman2010,B2004}. Usually all nodes are assumed to belong
to one population or class, and the interactions between two distinct populations
have been reported just recently \cite{zia2010, Stanley2010}.
The dynamical properties of two interacting populations, extroverts and
introverts were recently studied using dynamical network evolution model
 \cite{zia2010}. 
Buldyrev et al \cite{Stanley2010} developed a framework for understanding the
robustness of interacting networks. Using the generating functions method
they present the exact analytical solutions for the critical fraction of
nodes, which upon removal, will lead to a complete fragmentation into 
interdependent networks. 

The focus of the current approach in understanding the formation and evolution 
of terrorist 
networks is on middle-range perspective as opposed to the micro-level approach 
that considers individual terrorist and macro-level analysis of the root causes 
of terrorism \cite{sageman2004}. In particular, the interest is in placing 
the relationships between individuals in the context of (i) their interactions 
with each other, (ii) how they are influenced by ideas originating from their 
environment, (iii) their interactions with people and organizations outside of 
their group \cite{sageman2008}. Motivated by this description which conceptually 
refers to a complex system and because networks provide a fruitful framework to 
model complex systems \cite{Boccara}, we introduce a network model that
aims to describe the interactions between potentially terrorist and non-terrorist
populations.

Our goal is to present a model of a social network that contains two types
of agents which demonstrate different affinities in establishing connections
within their own population versus connections with the other population. We also 
aim that the model is simple enough to allow the derivation of 
approximate analytical expressions of the basic characteristics of the network.
The latter can be achieved 
in the general framework of rate equation theory in the mean field approximation
which has been introduced to study fundamental characteristics of growing 
network models \cite{BAJ99,Kertesz2003}. Within this framework we introduce 
the rate equations that are specific to our model, solve them and obtain 
analytical expressions that predict the growth dynamics of the degree of 
individual vertices. 

\begin{figure}
\noindent\includegraphics[width=26pc]{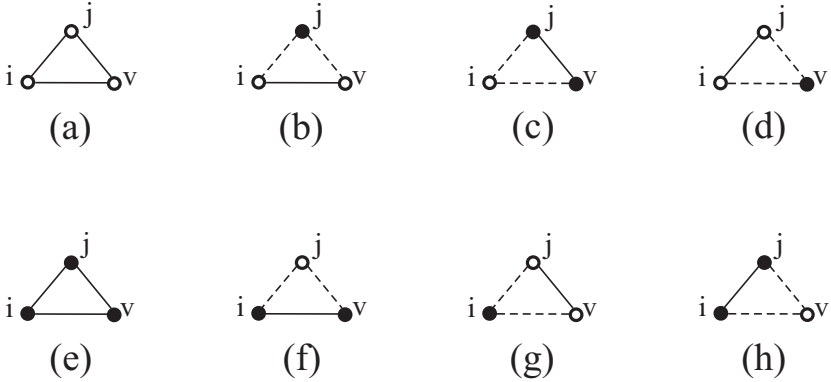}
\caption{Possible combinations of connections between a new node $v$, the
initial contact $j$ and the secondary contact $i$. 
(a)-(d) show the connections when the new node is an $N$ node. (e)-(h) show the
connections when the new node is a $V$ node. 
Empty symbols ($\circ$) mark an $N$ node while full symbols ($\bullet$)
mark a $V$ node. Full lines represent links between nodes
of the same type, e.g. friends, such as $NN$ or $VV$.
Dashed lines represent links between nodes of different types such as $NV$,
e.g. enemies.}
\label{fig1}
\end{figure}

While the probability to create initial connections is easily defined,
the probability for the secondary contacts between nodes is difficult to derive 
because of the interactions between the two types of nodes. As
a first approximation, we assume that the initial and secondary contacts 
form edges with the same probabilities.   
To include a more precise contribution of both
the initial and the secondary contacts in the rate equations we empirically obtain 
their functional dependences. This leads to an improved agreement between
analytical and numerical simulation results. 

From the functional dependence of the degree of 
a node as a function of time using the mean-field arguments, we derive
the degree distribution for each of the two types of nodes. We also derive 
analytical expressions of the structural, 
three-point correlations between nodes to study the clustering properties of
the networks. 

\section{Model}

The network models are broadly classified into two categories: the network evolution 
models in which addition of new edges depends on the local structure of 
the network, and nodal attribute models in which the existence of edges 
is determined solely by the attributes of the nodes 
(for review see \cite{toivonen2009}). The network evolution models 
can be further categorized into growing network evolution models and dynamical 
network evolution models. 
In the former the network growth starts with a small seed network and nodes 
and links are added according to specific rules until the network reaches a 
predetermined size. Dynamical network evolution models start with an empty 
network and edges are added and deleted according to specific rules until 
statistical properties of the network stabilize. 

In this paper, we propose a model that incorporates a growing network evolution 
process with nodal attributes which could be thought of as a new class of model. 
The model has five free parameters, three describing the growing network evolution
process and two describing nodal attributes. 
The parameters of the growing 
network evolution process are the number of nodes $\mathcal{N}$, the average number of 
nodes selected at random as initial contacts $m_r$, and the average number 
of nodes selected as secondary contacts $m_s$
among the neighbors of each initial contact \cite{toivonen2006}.
Two parameters quantify the type and amount 
of nodes, non-violent $N$ or potentially violent $V$ nodes and the type of 
interactions between them which are as follows:
(i) nodes are randomly marked as non-violent $N$ with probability 
$p_N$ and potentially violent $V$ with probability $p_V=1-p_N$; 
(ii) when establishing initial contacts, nodes connect 
with probability $p_s$ if the nodes are of the same type, 
such as $V$ with $V$ or $N$ with $N$, or with probability $p_d=1-p_s$ 
if the nodes are of different types such as $N$ with $V$ nodes.
The secondary contacts are established with nodes among the neighbors of
the initial contacts. The model combines the random attachment
of initial contacts with the implicit preferential attachment of the 
secondary contacts. In that, the model represents a generalization with 
two types of node attributes of growing models such as 
\cite{IJBC,Vazquez2003,toivonen2006}.

The definition of establishing edges can be thought of as creating links between
friends (solid lines in Fig.~\ref{fig1}) or between enemies (dashed lines).
By varying the value of $p_s$, we can generate different strengths of 
friendliness or animosity. 
The possible configurations of triads that arise in social networks 
in such a context are: (a) three friendly interactions; (b) one friendly 
and two unfriendly connections; (c) two friendly interactions and one 
unfriendly; (d) three unfriendly interactions \cite{newman2010}. According 
to the strong formulation of structural balance theory in social sciences,  
configurations (a) and (b) are considered stable while (c) and (d) are 
unstable and likely to break apart \cite{CH1956}. In their empirical large-scale 
verification of the long-standing structural balance theory, the authors of 
\cite{PNAS2010} find that the unstable triads, especially formation (c) 
are extremely underrepresented in an online social system in comparison 
to a null model. Our model produces correctly the stable configurations 
(a) (Fig.~\ref{fig1}a,e) and (b)  (Fig.~\ref{fig1}b-d,f-h) but cannot 
produce the unstable configurations (c) and (d) in accordance with the 
strong formulation of the structural balance theory  \cite{CH1956}.
 
The model algorithm consists of the following steps: 
(1) start with a seed network of $n_0$ connected nodes among which some are
$N$ and some are $V$, depending on $p_N$;
(2) at each time step add a new node, which has probability $p_N$ to be a 
non-violent and $p_V=1-p_N$ to be a violent node; 
(3) select on average $m_r\ge 1$ random nodes as initial contacts. 
The probability to connect the same type of nodes is $p_s$ 
while $p_d=1-p_s$ is the probability for initial contacts if they are of 
different types. 
(4) select on average $m_s\ge 0$ nodes among the neighbors of each initial
contact as secondary contacts. Connecting the new node with the secondary contacts 
is done without checking if it is the same type of node or not.
 There are two reasons for this choice. (i) Because the probability to 
establish inter-population connections ($p_s$) is higher than the probability 
to connect nodes intra-population ($p_d$), it is more likely that the first 
contact and its neighbors (potential secondary contacts) are of the same 
type than of different types. Therefore the secondary contacts will be more 
likely to be intra-population contacts even without explicitly modifying 
their probability to connect based on the type.  (ii) The secondary contacts 
are meant to mimic the `friend-of-a-friend' type of contacts in the real world, 
and we think that the implicit preferences given by the existing network 
connections would more accurately describe the nature of such contacts 
without including an explicit separate probability.
Apply steps (2) to (4)  until the network reaches the necessary size. 
 
\subsection{Rate Equations}

We start with constructing the rate equations that describe the change of 
the degree of a node on average during one time step of the network growth 
process for each of the non-violent $N$ and violent $V$ nodes. 
The degree of a node grows via two processes. One is the random attachment
of connecting a new node to $m_r$ nodes that are its initial contacts. 
The second process is when the new node is further connected to the $m_s$
nodes among the neighbors of the initial contacts. 
In the following we assume that the probability of this second process is 
linear with respect to the degree of the node which leads to implicit 
preferential attachment. The rate equations are:
\begin{eqnarray}
\frac{\partial k_{i(N)}}{\partial t}&=&
\frac{m_r}{tp_N}\left(p_Np_s+p_Vp_d\right) \nonumber \\
&+&m_rm_sp_N\left(\frac{p_sk_{is}}{\displaystyle\sum_{j,NN}k_j+
\displaystyle\sum_{j,NV}k_j}+
\frac{p_dk_{id}}{
\displaystyle\sum_{j,VV}k_j+\displaystyle\sum_{j,NV}k_j}\right) \nonumber\\
&+&m_rm_sp_V\left(\frac{p_sk_{id}}{\displaystyle\sum_{j,VV}k_j+
\displaystyle\sum_{j,NV}k_j}+
\frac{p_dk_{is}}{\displaystyle\sum_{j,NN}k_j+
\displaystyle\sum_{j,NV}k_j}\right) 
\label{N}
\end{eqnarray}
\begin{eqnarray}
\frac{\partial k_{i(V)}}{\partial t}&=&
\frac{m_r}{tp_V}\left(p_Vp_s+p_Np_d\right) \nonumber\\
&+&m_rm_sp_V\left(\frac{p_sk_{is}}{\displaystyle\sum_{j,VV}k_j+
\displaystyle\sum_{j,NV}k_j}+
\frac{p_dk_{id}}{
\displaystyle\sum_{j,NN}k_j+\displaystyle\sum_{j,NV}k_j}\right) \nonumber \\
&+&m_rm_sp_N\left(\frac{p_sk_{id}}{\displaystyle\sum_{j,NN}k_j+
\displaystyle\sum_{j,NV}k_j}+
\frac{p_dk_{is}}{\displaystyle\sum_{j,VV}k_j+
\displaystyle\sum_{j,NV}k_j}\right), 
\label{V}
\end{eqnarray}
where $k_i$ is the degree of node $i$ and we assumed that 
$k_{is}=k_ip_s$ and $k_{id}=k_ip_d$. 

All possible combinations of triads of a new node $v$, the 
initial contact $j$, and the secondary contact $i$ 
are schematically shown in Fig.~\ref{fig1}(a-h) and presented 
by the third through the sixth terms in 
Eq.~(\ref{N}) and Eq.~(\ref{V}) for an $N$ and $V$ node, respectively.

For example, the fifth term in Eq.~(\ref{N}) describes the rate of
change of the degree of vertex $i$ (which is an $N$ node) due to establishing 
the configuration of contacts shown in Fig.~\ref{fig1}c. 
The fifth term contains four factors. The first factor is the average number of 
secondary contacts which is $m_rm_s$. $p_V$ is the probability that the 
new node created at time step $t$ is a $V$ node.
$p_s$ is the probability that the newly created node connects to 
the initial contact which is a node of the same type.
Finally, $k_{id}/(\sum_{j,VV}k_j+\sum_{j,NV}k_j)$ is  the probability that 
the node selected for initial contact (node $j$ in Fig.~\ref{fig1}c) shares 
an edge with the node $i$.
This is a standard expression for preferential attachment except 
for the complications induced by having two distinct populations. 
The nominator $k_{id}$ is the degree of the $i$ node if we
count only the links to different types of nodes (in this case $V$ nodes).
The denominator is the sum of all possible links which the type of node 
selected for initial contact (in this case $V$) could have.
Assuming that the initial and secondary contacts are created with the same 
probabilities the nominator would be
equal to $k_{id}=k_ip_d$ and we will approximate it in this way and
the denominator would be equal to $(p_Vp_s+p_d)2\,m_r(1+m_s)t$.
We will use this expression as a first approximation and we will also
empirically derive functional dependences of the denominator and 
compare the results.
 
Note that the number of edges, multiplied by two, that exist between 
$N$ and $N$ nodes is equal to
$\sum_{j,NN}k_j$ and include both edges created as initial 
contacts and edges created as secondary contacts. The same is true for 
the number of edges between $V$ and $V$ nodes which is  
$\sum_{j,VV}k_j$, and the number of edges between $N$ and $V$ nodes 
which is $\sum_{j,NV}k_j$ in Eqs.~(\ref{N}) and (\ref{V}). 
We know that the probabilities for creating edges as initial contacts 
are $p_s$ or $p_d$ if between same type of nodes or different types of nodes,
respectively. We do not know, however, what these probabilities are when the 
edges represent connections to secondary contacts. This is the reason to express the 
respective summations by the following relations:
\begin{subequations}
\label{sms}
\begin{align}
\displaystyle\sum_{j,NN}k_j&=g(\rho_N(p_N,p_s))2\,m_r(1+m_s)t \label{sNN} \\
\displaystyle\sum_{j,VV}k_j&=h(\rho_V(p_V,p_s))2\,m_r(1+m_s)t \label{sVV}\\
\displaystyle\sum_{j,NV}k_j&=q(\rho_d(p_d))2\,m_r(1+m_s)t. \label{sNV}
\end{align}
\end{subequations}
Eqs.~(\ref{sms}) contain the term $2\,m_r(1+m_s)t$ because there are  
$\sim t$ vertices at time $t$ and $m_r(1+m_s)$ is the average initial degree
of a vertex. $g(\rho_N(p_N,p_s))$ in Eq.~(\ref{sNN}) is the probability that 
the edge is between $N$ nodes. $h(\rho_V(p_V,p_s))$ Eq.~(\ref{sVV}) is the 
probability that the edge is between $V$ nodes, and $q(\rho_d(p_d))$ 
Eq.~(\ref{sNV}) is the probability that the edge is between
nodes of different types. Functions $g$, $h$, and $q$ contain edges established
due to both initial and secondary contacts, whose contributions cannot be 
separated and derived analytically. Therefore, we will obtain these functional 
dependences
through empirical considerations. If we assume that the edges to secondary contacts
are established with the same probabilities as the edges to initial contacts, then
the relations would have been:
\begin{subequations}
\label{simple}
\begin{align}
g(\rho_N(p_N,p_s))&=p_Np_s \label{mNN}\\
h(\rho_V(p_V,p_s))&=p_Vp_s \label{mVV}\\
q(\rho_d(p_d))&=p_d. \label{mNV}
\end{align}
\end{subequations} 

\subsection{Solutions of the Rate Equations}

After separating the variables and integrating the rate equation of the 
degree $k_{i(N)}$ of $N$ node Eq.(\ref{N}) from $k_{init}$ 
to $k_i$ and from $t_i$ to $t$ we obtain the following expression for the
degree as a function of time $k_i(t)$ 
\begin{equation}
k_{i(N)}(t)=G_3\left(\frac{t}{t_i}\right)^{1/G_1}-G_2,
\label{kin}
\end{equation}
where 
\begin{subequations}
\label{knt}
\begin{align}
G_1&=C\left(\frac{p_Np_s^2+p_Vp_sp_d}{g(\rho_N)+q(\rho_d)}+
\frac{p_Np_d^2+p_Vp_sp_d}{h(\rho_V)+q(\rho_d)}\right)^{-1} \label{knt1}\\
G_2&=A_NG_1 \label{knt2}\\
G_3&=G_2+k_{init} \label{knt3} \\
C&=2(1+m_s)/m_s \label{knt4} \\
A_N&=m_r\left(p_Np_s+p_Vp_d\right)/p_N \label{knt5}\\ 
k_{init}&=m_r(1+m_s). \label{knt6}
\end{align}
\end{subequations}
Integrating the rate equation of the degree $k_i$ of $V$ nodes, Eq.(\ref{V})
produces the time dependence of the degree of any $V$ node
\begin{equation}
k_{i(V)}(t)=H_3\left(\frac{t}{t_i}\right)^{1/H_1}-H_2,
\label{kiv}
\end{equation}
where 
\begin{subequations}
\label{kvt}
\begin{align}
H_1&=C\left(\frac{p_Vp_s^2+p_Np_sp_d}{h(\rho_V)+q(\rho_d)}+
\frac{p_Vp_d^2+p_Np_sp_d}{g(\rho_N)+q(\rho_d)}\right)^{-1} \label{kvt1} \\
H_2&=A_VH_1 \label{kvt2} \\
H_3&=H_2+k_{init} \label{kvt3} \\
A_V&=m_r\left(p_Vp_s+p_Np_d\right)/p_V. \label{kvt4}
\end{align}
\end{subequations}
If we assume that the edges to secondary contacts
are established with the same probabilities as the edges to initial contacts, then
the solution of the rate equation Eq.~(\ref{N}) will be the 
following for an $N$ node 
\begin{equation}
k_{i(N)}(t)=D_3\left(\frac{t}{t_i}\right)^{1/D_1}-D_2,
\label{kins}
\end{equation}
where 
\begin{subequations}
\label{knts}
\begin{align}
D_1&=C\left(\frac{p_Np_s^2+p_Vp_sp_d}{p_np_s+p_d}+
\frac{p_Np_d^2+p_Vp_sp_d}{p_vp_s+p_d}\right)^{-1} \label{knts1}\\
D_2&=A_ND_1 \label{knts2}\\
D_3&=D_2+k_{init}, \label{knts3} 
\end{align}
\end{subequations}
and for a $V$ node:
\begin{equation}
k_{i(V)}(t)=E_3\left(\frac{t}{t_i}\right)^{1/E_1}-E_2,
\label{kivs}
\end{equation}
where 
\begin{subequations}
\label{kvts}
\begin{align}
E_1&=C\left(\frac{p_Vp_s^2+p_Np_sp_d}{p_vp_s+p_d}+
\frac{p_Vp_d^2+p_Np_sp_d}{p_np_s+p_d}\right)^{-1} \label{kvts1} \\
E_2&=A_VE_1 \label{kvts2} \\
E_3&=E_2+k_{init}, \label{kvts3} 
\end{align}
\end{subequations}
 after making use of Eqs.~(\ref{simple}).

\subsection{Degree distribution}

In the mean field approximation, the degree $k_i(t)$ of a node $i$ evolves 
with time $t$ strictly monotonically after the node was added to the network 
at time $t_i$. Therefore, the nodes added to the network more recently 
will have on average lower degree 
than those added to the network earlier. Assuming that we add nodes to 
the network at equal intervals, the probability 
density of $t_i$ is $1/t$. Using the properties of cumulative 
probability distribution function, we can write that the probability
of a node to have degree $\tilde{k}\le k_i$ is equal to the
probability that the node has been added to the network at 
time $\tilde{t}\ge t_i$
\begin{equation}
F(k)=P(\tilde{k}\le k_i)=P(\tilde{t}\ge t_i)=\frac{t-t_i}{t}.
\label{cdf}
\end{equation}
We can derive the probability density function of $N$ nodes, $P_N(k)$ 
by obtaining an expression for $t_i^{(N)}$ from Eq.~(\ref{kin}), then  
replacing it in Eq.~(\ref{cdf}) and differentiating the resultant equation with 
respect to $k_{i}$, which is $P(k)=\partial F/\partial k$. The result is:
\begin{equation}
P_N(k)=G_1(G_2+k)^{-G_1-1}G_3^{G_1}.
\label{pn}
\end{equation}
Similarly, we can derive the probability density function of $V$ nodes, 
$P_V(k)$ by obtaining an expression for $t_i^{(V)}$ from Eq.~(\ref{kiv}), then  
replacing it in Eq.~(\ref{cdf}) and differentiating the resultant equation with 
respect to $k_{i}$ to obtain:
\begin{equation}
P_V(k)=H_1(H_2+k)^{-H_1-1}H_3^{H_1}.
\label{pv}
\end{equation}
If we use Eqs.~(\ref{kins}) and (\ref{kivs}) to derive expressions for 
$t_i^{(N)}$ and $t_i^{(V)}$, respectively, then the degree distributions are
presented by
\begin{equation}
P_N(k)=D_1(D_2+k)^{-D_1-1}D_3^{D_1}
\label{pns}
\end{equation}
\begin{equation}
P_V(k)=E_1(E_2+k)^{-E_1-1}E_3^{E_1}.
\label{pvs}
\end{equation}
It should be noted that $k_i$ and all quantities are expectation values
and can be compared to simulation results which are assemble averages. 
The analytical results converge to those reported in Ref. \cite{toivonen2006} 
in the limit of one population which means $p_N=1$, $p_V=0$, $p_s=1$, 
and $p_d=0$.

\subsection{Clustering Characteristics}

The dependence of the clustering coefficient as a function of the degree of 
a node can be derived using the rate equation method \cite{BAJ99,Kertesz2003}.
The number of triangles $E_{i(N)}$ ($E_{i(V)}$) around a node if $i$ is an
$N$ ($V$) node is changing with time following two processes. The first 
process is when node $i$ is selected as one of the initial contacts with 
probability $A_N/t$ ($A_V/t$) and the new node links to some of its neighbors
which are $m_s$ on average. The second process is when node $i$ is selected 
as a secondary contact and a triangle is formed between the new node, the 
initial contact, and the secondary contact. It is possible that two neighboring
initial contacts and the new node form a triangle, but the contribution
of this process is negligible. The rate equation for the number of 
connections between the nearest neighbors of a node of degree $k_{iN}$
($k_{iV}$) is given by
\begin{eqnarray}
\frac{\partial E_{i(N)}}{\partial t}&=&
\frac{m_rm_s}{tp_N}\left(p_Np_s+p_Vp_d\right) \nonumber \\
&+&m_rm_sp_N\left(\frac{p_sk_{is}}{\displaystyle\sum_{j,NN}k_j+
\displaystyle\sum_{j,NV}k_j}+
\frac{p_dk_{id}}{
\displaystyle\sum_{j,VV}k_j+\displaystyle\sum_{j,NV}k_j}\right) \nonumber\\
&+&m_rm_sp_V\left(\frac{p_sk_{id}}{\displaystyle\sum_{j,VV}k_j+
\displaystyle\sum_{j,NV}k_j}+
\frac{p_dk_{is}}{\displaystyle\sum_{j,NN}k_j+
\displaystyle\sum_{j,NV}k_j}\right) 
\label{EN}
\end{eqnarray}
\begin{eqnarray}
\frac{\partial E_{i(V)}}{\partial t}&=&
\frac{m_rm_s}{tp_V}\left(p_Vp_s+p_Np_d\right) \nonumber \\
&+&m_rm_sp_V\left(\frac{p_sk_{is}}{\displaystyle\sum_{j,VV}k_j+
\displaystyle\sum_{j,NV}k_j}+
\frac{p_dk_{id}}{
\displaystyle\sum_{j,NN}k_j+\displaystyle\sum_{j,NV}k_j}\right) \nonumber \\
&+&m_rm_sp_N\left(\frac{p_sk_{id}}{\displaystyle\sum_{j,NN}k_j+
\displaystyle\sum_{j,NV}k_j}+
\frac{p_dk_{is}}{\displaystyle\sum_{j,VV}k_j+
\displaystyle\sum_{j,NV}k_j}\right),
\label{EV}
\end{eqnarray}
respectively. After some algebra and using Eq.~(\ref{N}) if $i$ is an $N$
node and Eq.~(\ref{V}) if $i$ is a $V$ node we obtain 
\begin{equation}
\frac{\partial E_{i(N)}}{\partial t}=\frac{\partial k_{i(N)}}{\partial t}+
\frac{(m_s-1)A_N}{t}
\label{EN1}
\end{equation}
and
\begin{equation}
\frac{\partial E_{i(V)}}{\partial t}=\frac{\partial k_{i(V)}}{\partial t}+
\frac{(m_s-1)A_V}{t}.
\label{EV1}
\end{equation}
After integrating both sides of Eqs.~(\ref{EN1}) and (\ref{EV1})
with respect to $t$, using
the initial condition $E_{i(N)}(k_{init},t_i)=E_{i(V)}(k_{init},t_i)=m_rm_s$,
and $k_{init}$ given by Eq.~(\ref{knt6}), we obtain the expressions for 
the change with time of the number of connections between the nearest neighbors 
of a node of degree $k_{iN}$ 
\begin{equation}
E_{i(N)}=k_{i}-m_r+
(m_s-1)A_N\ln\left(\frac{t}{t_i}\right)
\label{EN2}
\end{equation}
if $i$ is an $N$ node and of a node of degree $k_{iV}$
\begin{equation}
E_{i(V)}=k_{i}-m_r+
(m_s-1)A_V\ln\left(\frac{t}{t_i}\right)
\label{EV2}
\end{equation}
if $i$ is a $V$ node. We use Eqs.~(\ref{pn}) and (\ref{pv}) to obtain expressions
for $\ln(t/t_i)$ and insert them in Eqs.~(\ref{EN2}) and (\ref{EV2}), respectively. 
Finally, the degree-dependent clustering coefficient $C(k)$ as a function 
of the degree $k$ of the node  which is also referred as the clustering 
spectrum, is given by
\begin{equation}
C_N(k)=2\frac{k-m_r+
(m_s-1)A_NG_1\ln\left[\left(k+G_2\right)/G_3\right]}
{k(k-1)}
\label{ckN}
\end{equation}
for an $N$ node and by
\begin{equation}
C_V(k)=2\frac{k-m_r+
(m_s-1)A_VH_1\ln\left[\left(k+H_2\right)/H_3\right]}
{k(k-1)}
\label{ckV}
\end{equation}
for a $V$ node, where we make use of $C(k)=2E(k)/[k(k-1)]$, which 
defines the clustering coefficient of a vertex as the ratio of the
total number of existing connections between all $k$ of its neighbors
and the number $k(k-1)/2$ of all possible connections between them.
The degree-dependent clustering coefficient $C(k)$ defines the local
clustering properties of the network.

The global clustering characteristics of a network include
the mean clustering coefficient $\bar{C}$ as averaged over the vertex degree,
the mean clustering $<C>=1/n\sum_iC_i$ as averaged over the nodes of the network
(where $C_i$ is the clustering coefficient of node $i$),
and the so-called transitivity $C$ \cite{newman2010}.
Making use of degree-dependent local clustering coefficient 
(Eqs.~(\ref{ckN}) and (\ref{ckV})) and the degree distribution 
(Eqs.~(\ref{pns}) and (\ref{pvs}))
for $N$ or $V$ node one can define the respective mean clustering coefficient
as:
\begin{equation}
\bar{C}=\sum_k P(k)C(k).
\label{bc}
\end{equation}
Transitivity is a measure of the ratio of the total number of loops of 
length three in a graph to the total number of connected triples and is 
defined as \cite{newman2010,dorogovtsev2004}
\begin{equation}
C=\frac{\sum_k P(k)E(k)}{\sum_kP(k)k(k-1)/2}.
\label{T}
\end{equation}
The mean clustering coefficient $\bar{C}$ and the transitivity $C$ assess in
a different manner the clustering properties of a graph. In real networks they 
could have very different values for the same network \cite{newmanSIAM}. 

\section{Comparison between Theory and Simulation Results}

We compare three outputs: the analytical derivation of degree 
distribution $P(k)$ (Eqs.~(\ref{pns}) and (\ref{pvs})) that was obtained assuming
that the links to secondary contacts are established with the same probabilities
as the links to initial contacts Eq.~(\ref{simple}), the derivation of 
$P(k)$ (Eqs.~(\ref{pn}) and (\ref{pv})) obtained by using functional 
dependences Eq.~(\ref{sms}), and the numerical simulation results. 

\begin{figure}
\noindent\includegraphics[width=15pc]{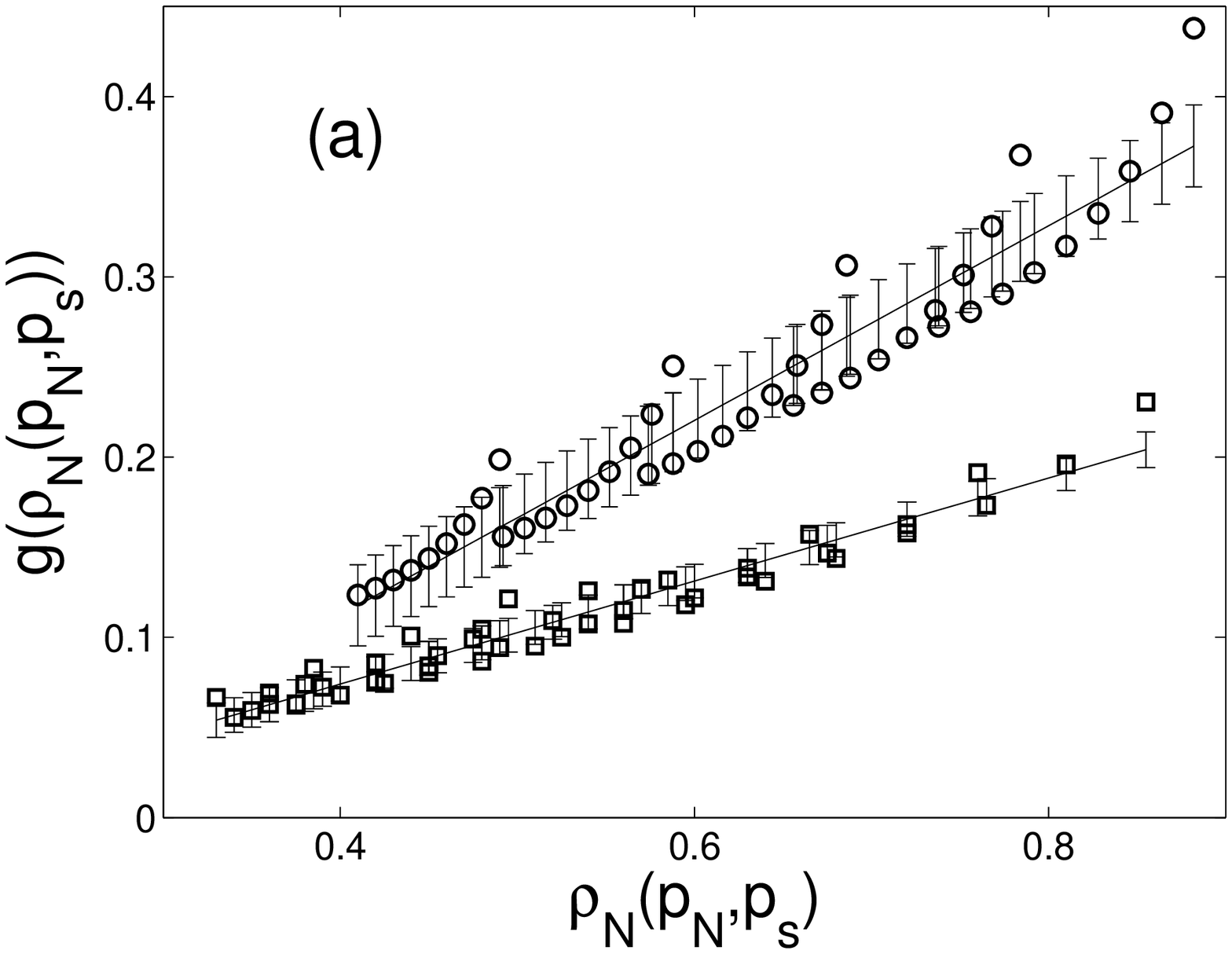}
\hfill
\noindent\includegraphics[width=15pc]{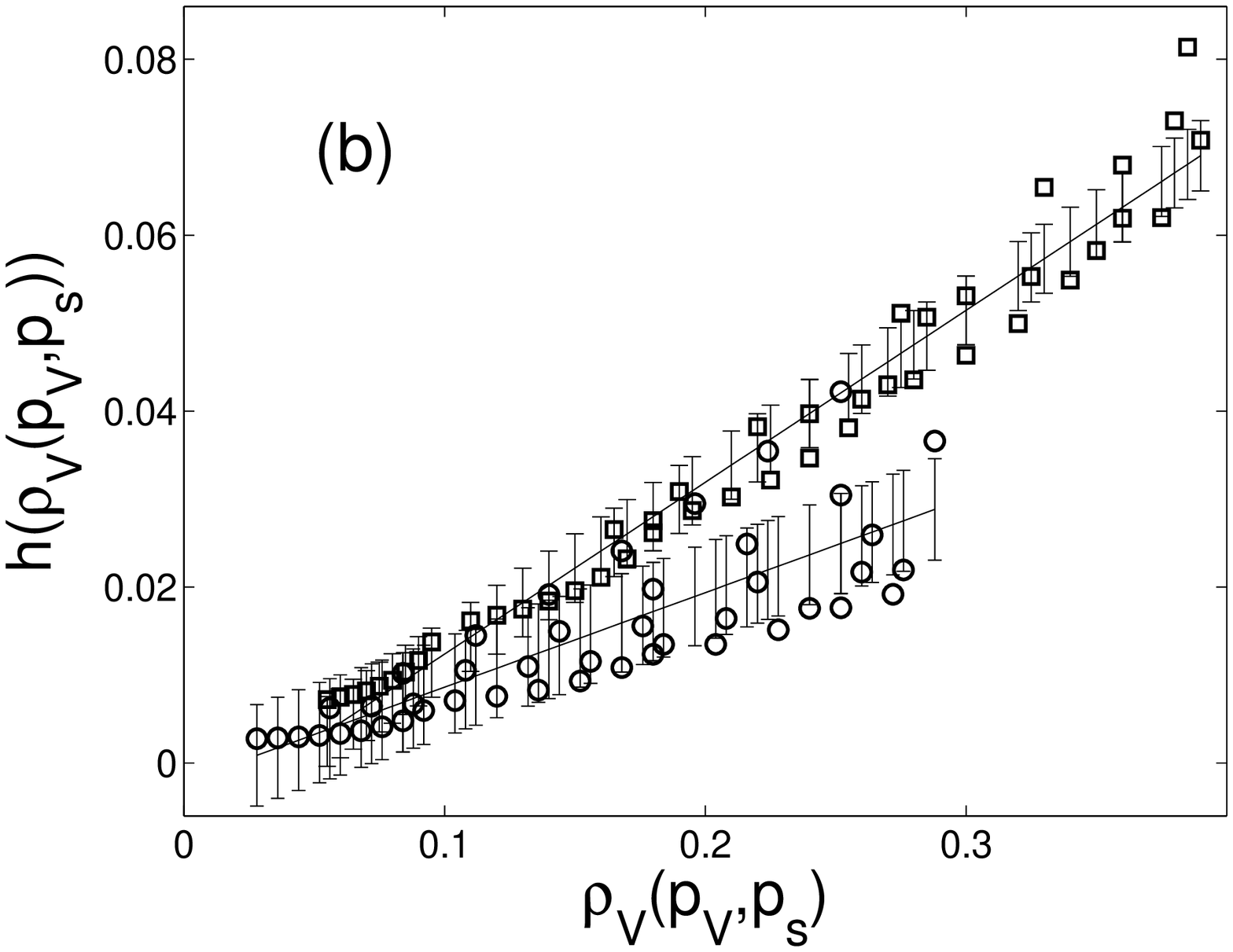}
\vfill
\noindent\includegraphics[width=15pc]{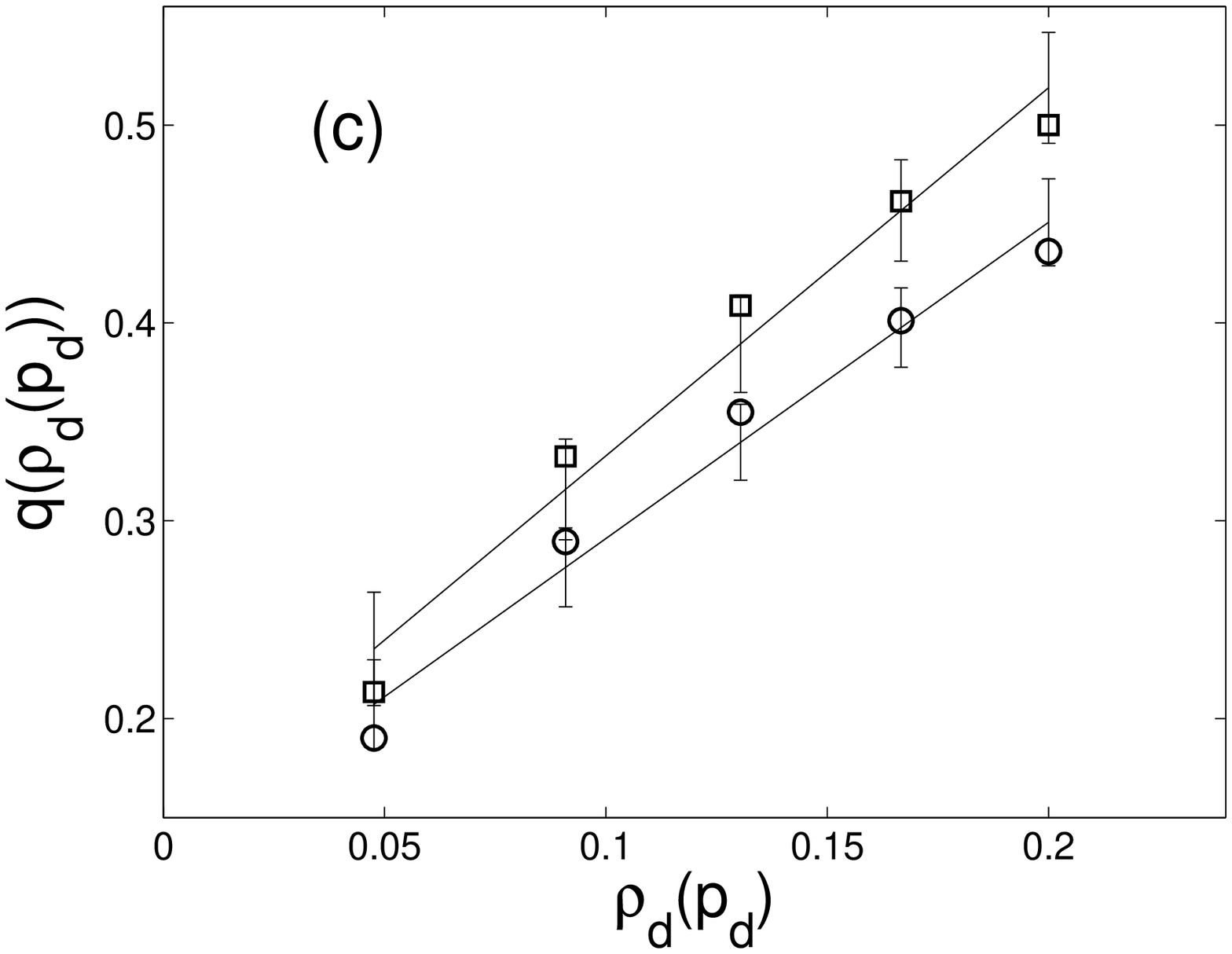}
\caption{(a) Probability to establish both initial and secondary contacts
between $N$ nodes $g(\rho_N(p_N,p_s))$ as a function of 
$\rho_N(p_N,p_s)=p_N(c_{p_N}+p_s(1-c_{p_N}))$, where $c_{p_N}=0.8$
for $p_N=0.8$ ($\bigcirc$) and $c_{p_N}=0.5$ for $p_N=0.5$ ($\square$).
(b) Probability to establish both initial and secondary contacts
between $V$ nodes $h(\rho_V(p_V, p_s))$
as a function of $\rho_V(p_V,p_s)=p_V(c_{p_V}+p_s(1-c_{p_V}))$, where 
$c_{p_V}=1-c_{p_N}=0.2$ for $p_V=0.2$ ($\bigcirc$) and $c_{p_V}=0.5$ 
for $p_V=0.5$ ($\square$). 
(c) Probability to establish both initial and secondary contacts
between different types of nodes as a function of $\rho_d(p_d)=p_d/(p_d+2)$  
for $p_d\in[0.1, 0.5]$ at fixed $p_N=0.8$ ($\bigcirc$), 
and fixed $p_N=0.5$ ($\square$), respectively. 
Values on y-axis in (a) represent matrix multiplication of
the vector of $g(\rho_N(p_N,p_s))$ as a function 
of $p_N\in[0.5, 0.9]$ at fixed $p_s=0.7$ ($\bigcirc$) ($p_s=0.5$ ($\square$))
multiplied by the vector of normalized number of $NN$ edges as a function 
of $p_s\in[0.5, 0.9]$ at fixed $p_N=0.8$ ($\bigcirc$) ($p_N=0.5$ ($\square$)).
Values on y-axis in (b) represent matrix multiplication of
the vector of $h(\rho_V(p_V, p_s))$ as a function 
of $p_V\in[0.1, 0.5]$ at fixed $p_s=0.7$ ($\bigcirc$) ($p_s=0.5$ ($\square$))
multiplied by the vector of normalized number of $VV$ edges as a function 
of $p_s\in[0.5, 0.9]$ at fixed $p_V=0.2$ ($\bigcirc$) ($p_V=0.5$ ($\square$)).
Case I: one node as initial contact and two nodes as
secondary contacts.
}
\label{fig2}
\end{figure}

\subsection{Functional dependence}

First, let us focus on the functional dependence of the probability that the 
edge is between $N$ nodes $g(\rho_N)$ on $\rho_N$ (Eq.~(\ref{sNN})), which 
represents the combined probabilities that the edge is between same type of 
nodes $p_s$ and the probability that they are $N$ nodes. We empirically estimate 
$g(\rho_N(p_N,p_s))$ by calculating the number of $NN$ edges, multiplying it 
by two and dividing it by the average number of nodes at time $t$ which is 
$2m_r(1+m_s)t$. We obtain the functional dependence  of empirically estimated 
$g(\rho_N)$ by a matrix multiplication of two vectors; one is 
the vector of $g(\rho_N)$ values as a function of $p_N$ at fixed 
$p_s=0.7$ and the other is the vector of $g(\rho_N)$ values as a 
function of $p_s$ at fixed $p_N=0.8$. Next, we aim to construct a function
of $p_N$ and $p_s$, $\rho_N(p_N,p_s)$ such that the empirically estimated 
$g(\rho_N(p_N,p_s))$ which express the probabilities for both initial
and secondary contacts is a linear function of $\rho_N$. We plot the result 
for Case I (one initial contact and two secondary contacts) in
Fig.~\ref{fig2}a for
$p_N\in[0.5, 0.9]$ at fixed values of $p_s=0.7$ and for $p_s\in[0.5, 0.9]$ 
at fixed values of $p_N=0.8$ (circles $\bigcirc$). 
Squares ($\square$) Fig.~\ref{fig2}a mark 
results for $p_N\in[0.5, 0.9]$ at fixed values of $p_s=0.5$ and 
for $p_s\in[0.5, 0.9]$ at fixed values of $p_N=0.5$.
We obtain that the combined probability $\rho_N(p_N,p_s)$ of the form
\begin{equation}
\rho_N(p_N,p_s)=p_N(c_{p_N}+p_s(1-c_{p_N})),
\end{equation} 
produces the following least-squares linear fit 
\begin{subequations}
\label{parN}
\begin{align}
g(\rho_N)=0.54\rho_N-0.10\pm\Delta_N \qquad (\bigcirc) \label{pN1}\\
g(\rho_N)=0.29\rho_N-0.04\pm\Delta_V \qquad (\square) \label{pN2}
\end{align}
\end{subequations} 
for $c_{p_N}=0.8$ ($\bigcirc$) and $c_{p_N}=0.5$ ($\square$), respectively.

The prediction error estimate was generated for $g(\rho_N)$ and found to be  
$\Delta_N=0.0221$ (for $\bigcirc$) and $\Delta_N=0.0095$ and
(for $\square$) which allows us to obtain a range of values for $g(\rho_N)$
limited by $g(\rho_N)\pm\Delta_N$. We use these functional dependences
within their range to express $g(\rho_N(p_N,p_s))$
in the solutions of the rate equations Eqs.~(\ref{kin}) and (\ref{kiv}) 
and in the expression of respective degree distributions and clustering
coefficients. 

Applying the same reasoning, we obtain the combined probability
\linebreak $h(\rho_V(p_V, p_s))$ as a function of $\rho_V$ which represents 
both the initial and secondary contacts. Results are shown 
in Fig.~\ref{fig2}b for $p_V\in[0.1, 0.5]$ at fixed $p_s=0.7$ and 
$p_s\in[0.5, 0.9]$ at fixed $p_V=0.2$ ($\bigcirc$). Squares ($\square$) 
mark results for $p_V\in[0.1, 0.5]$ at fixed $p_s=0.5$ and 
$p_s\in[0.5, 0.9]$ at fixed $p_V=0.5$.
For an argument of the form
\begin{equation}
\rho_V(p_V,p_s)=p_V(c_{p_V}+p_s(1-c_{p_V})),
\end{equation}
the least-square fit produces 
\begin{subequations}
\label{parV}
\begin{align}
h(\rho_V)=0.11\rho_V\pm\Delta_V \qquad (\bigcirc)\\ 
h(\rho_V)=0.20\rho_V-0.01\pm\Delta_V \qquad (\square) 
\end{align}
\end{subequations} 
for $c_{p_V}=1-c_{p_N}=0.2$ ($\bigcirc$) and $c_{p_V}=0.5$ ($\square$),
respectively. 
The prediction error estimate defines the range of values for 
$h(\rho_V)\pm\Delta_V$, where $\Delta_V=0.0056$ (for $\bigcirc$) and
$\Delta_V=0.0039$ (for $\square$).

We obtain the probability  $q(\rho_d(p_d))$ for creating a link between 
different type of nodes both as initial and as secondary contacts to be
\begin{subequations}
\label{pard}
\begin{align}
q(\rho_d)=1.60\rho_d+0.1\pm\Delta_d \qquad (\bigcirc) \\
q(\rho_d)=1.86\rho_d+0.15\pm\Delta_d \qquad (\square).
\end{align}
\end{subequations} 
The dependence of $q(\rho_d(p_d))$ as a function of $\rho_d$ is shown 
in Fig.~\ref{fig2}c for $p_d\in[0.5, 0.9]$ at fixed $p_N=0.8$ ($\bigcirc$) 
and $p_N=0.5$ ($\square$). 
We applied the linear least-square fit for an argument of the form 
\begin{equation}
\rho_d(p_d)=p_d/(p_d+2).
\end{equation}
The prediction error estimate is obtained to be $\Delta_d=0.0207$
(for $\bigcirc$) and $\Delta_d=0.0265$ (for $\square$).

We use the above parameterization procedures of the probabilities to establish
both initial and secondary contacts between $N$ nodes $g(\rho_N(p_N,p_s))$,
between $V$ nodes $h(\rho_V(p_V, p_s))$, and between different types of nodes
$q(\rho_d(p_d))$ in obtaining the degree distribution (Eqs.~(\ref{pn}) and 
(\ref{pv})) and clustering coefficient (Eqs.~(\ref{ckN}) and (\ref{ckV}))
for each of the cases considered below. 

\begin{figure}
\noindent\includegraphics[width=14pc]{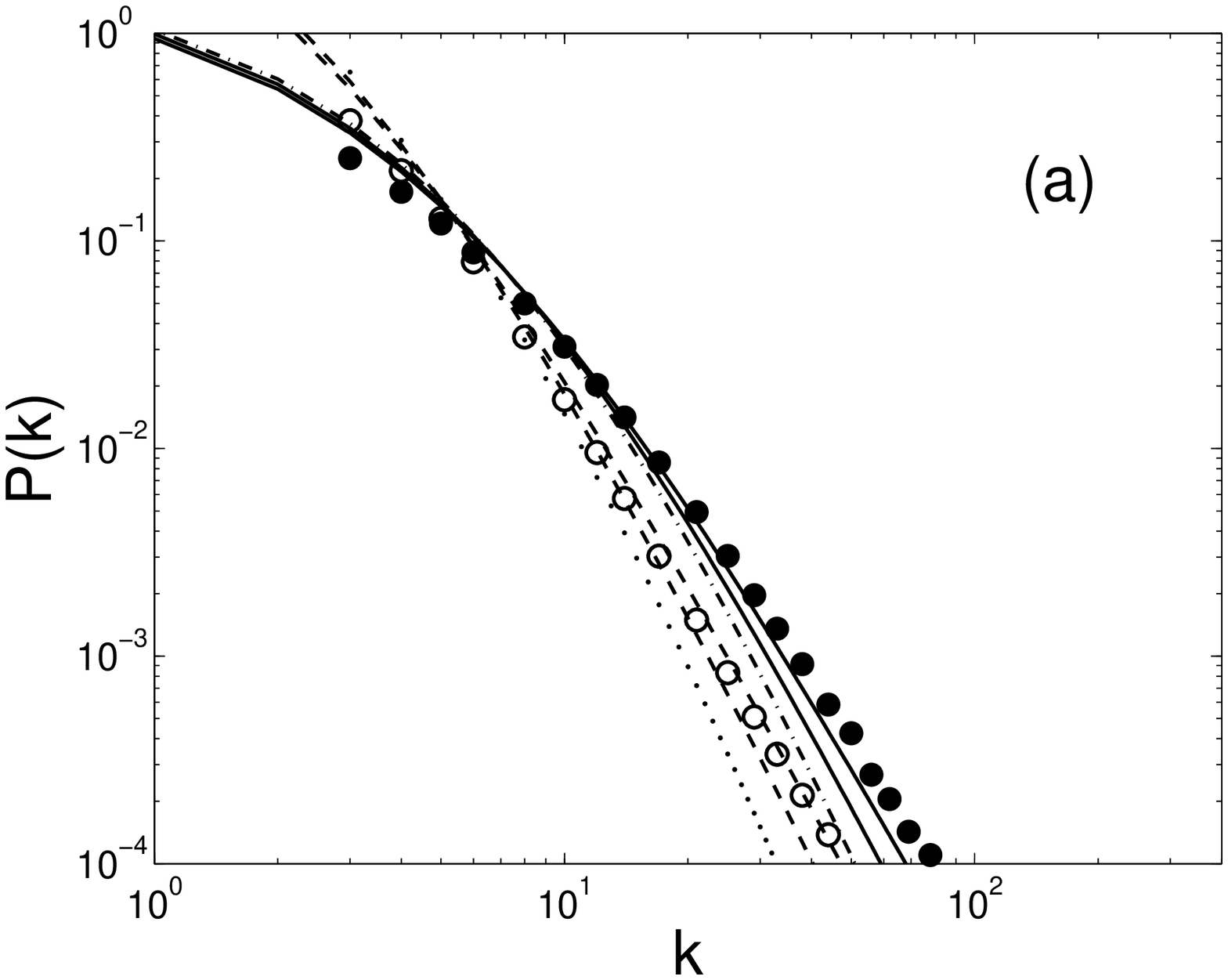}
\hfill
\noindent\includegraphics[width=14pc]{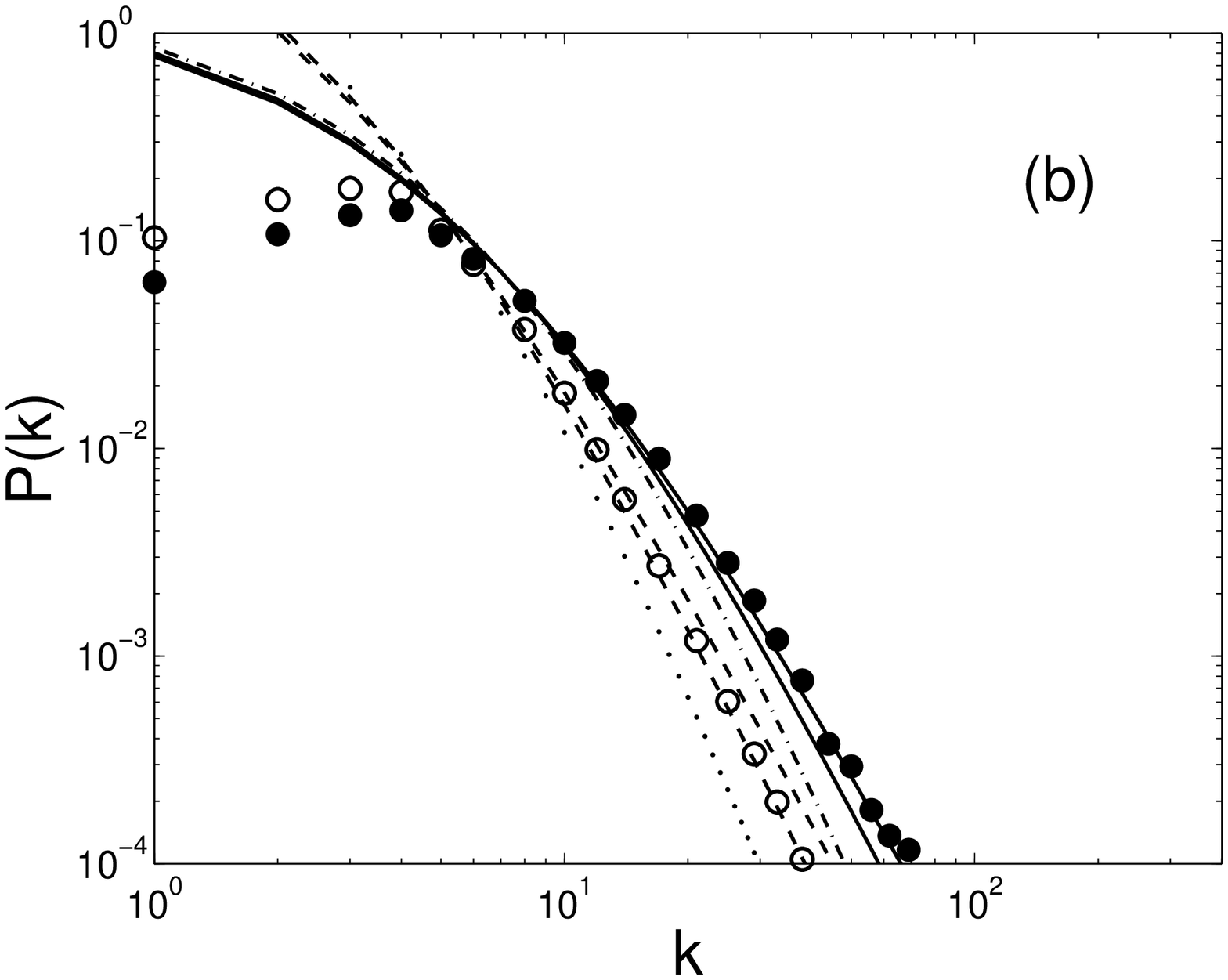}
\vfill
\noindent\includegraphics[width=14pc]{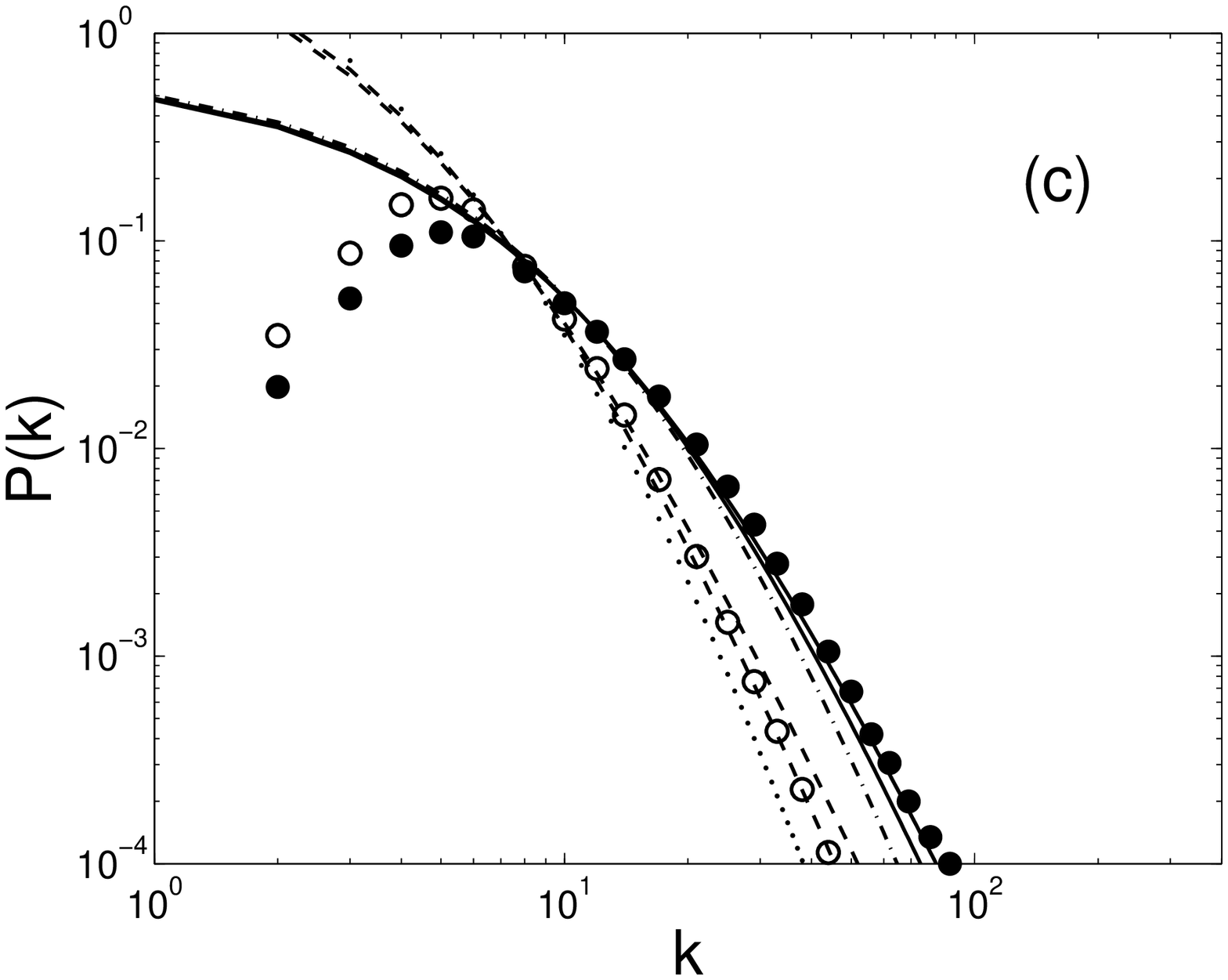}
\caption{Degree distribution for two populations consisting of 80\% $N$ 
nodes ($\circ$) $p_N=0.8$ and 20\% $V$ nodes 
$p_V=0.2$ ($\bullet$). Edges are formed with probability $p_s=0.7$ if
they connect $N$ with $N$ or $V$ with $V$ nodes and with probability
$p_d=0.3$ if they connect $N$ with $V$ nodes. 
Symbols represent results from numerical simulations done on a network with
$10^5$ agents and averaged over 100 runs.
Lines represent the results of
degree distribution $P(k)$ given by Eq.~(\ref{pn}) for $N$ nodes (solid) and
by Eq.~(\ref{pv}) for $V$ nodes (dashed) using functional dependences
shown in Fig.~\ref{fig2} to obtain functions $\rho_N\pm\Delta_N$, $\rho_V\pm\Delta_V$,
and $\rho_d\pm\Delta_d$. The set of two lines correspond to using $+\Delta_{\nu}$ or
$-\Delta_{\nu}$ for each of $\rho_{\nu}$, where $\nu=N, V, d$.
The dash-dotted line represents analytical distribution 
$P(k)$ obtained using Eq.~(\ref{pns})  for $N$ nodes. 
The dotted line represents analytical distribution 
$P(k)$ obtained using Eq.~(\ref{pvs}) for $V$ nodes.
(a) Case I: one node as initial contact and two nodes as secondary contacts; 
(b) Case II: one node as initial contact with probability 0.9 and
two nodes as initial contacts with probability 0.1; the number of nodes as
secondary contacts is from uniform distribution $U[0, 3]$;
(c) Case III: two nodes as initial contacts; the number of nodes as 
secondary contacts is from uniform distribution $U[0, 2]$.}
\label{fig3}
\end{figure}

\subsection{Degree distribution}

The analytical distribution $P(k)$ is
given by Eq.~(\ref{pns}) for $N$ nodes and by Eq.~(\ref{pvs}) for $V$ nodes
and plotted in Fig. \ref{fig3} with dash-dotted and dotted lines, respectively.
The degree distribution $P(k)$ obtained using functional dependences
is given by Eq.~(\ref{pn}) for $N$ nodes and
by Eq.~(\ref{pv}) for $V$ nodes and plotted in Fig. \ref{fig3} with solid and
dashed lines, respectively. 
Simulations are conducted on a network with $10^5$ nodes 
starting with a seed network of 8 nodes 
and are averaged over a 100 runs. All three cases considered are for value of the 
probability to create an $N$ node $p_N=0.8$ and for the value of the probability 
to establish a link between the same type of nodes $p_s=0.7$. 
To touch upon the versatility of the model we consider three different cases. 
They are Case I: one node as initial contact $m_r=1$ and 
two nodes as secondary contacts $m_s=2$ (Fig.~\ref{fig3}a); Case II: one node 
as initial contact with probability 0.9 and two nodes as initial contacts 
with probability 0.1, which gives $m_r=1.1$; the number of nodes as
secondary contacts is from uniform distribution $U[0, 3]$ and therefore, 
$m_s=1.5$ (Fig.~\ref{fig3}b); 
Case III: two nodes as initial contacts $m_r=2$; the number of nodes as 
secondary contacts is from uniform distribution $U[0, 2]$, $m_s=1$ 
(Fig.~\ref{fig3}c).
Results demonstrate that for all cases considered the simulations 
compare relatively well with the analytical derivation of the degree 
distribution even though 
using functional dependences in $P(k)$ derivation improves
the agreement within the limits of the simulations. 

\begin{figure}
\noindent\includegraphics[width=14pc]{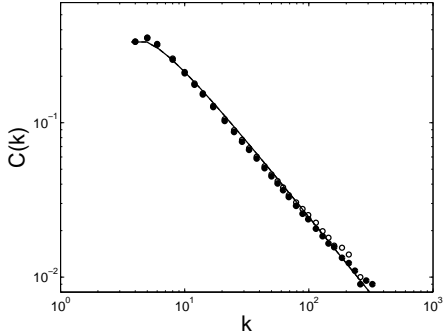}
\caption{Clustering coefficient as a function of the degree of the node
for $N$ nodes ($\circ$) and $V$ nodes ($\bullet$) for $p_N=0.8$
and $p_s=0.7$. Symbols represent results from numerical simulations. 
Lines (indistinguishable) represent the results of theoretical 
derivation of clustering 
coefficient as a function of the degree of the node using Eq.~(\ref{ckN})
for $N$ nodes and Eq.~(\ref{ckV}) for $V$ nodes. 
Case III: two nodes as initial contacts; the number of nodes as
secondary contacts is from uniform distribution $U[0, 2]$.}
\label{fig4}
\end{figure}

\subsection{Clustering}

The simulation results for the clustering coefficient as a function
of the degree of the node $C(k)$ for $N$ and $V$ nodes for Case III
are plotted in Fig.~\ref{fig4} with 
empty and full circles, respectively. 
The analytical solution for clustering coefficient $C(k)$ using Eq.~(\ref{ckN})
for $N$ nodes and Eq.~(\ref{ckV}) for $V$ nodes are plotted with lines
which coincide with each other. A clear $C(k)\sim 1/k$ trend is observed
which indicates the hierarchy in the system.  
 
\begin{figure}
\noindent\includegraphics[width=15pc]{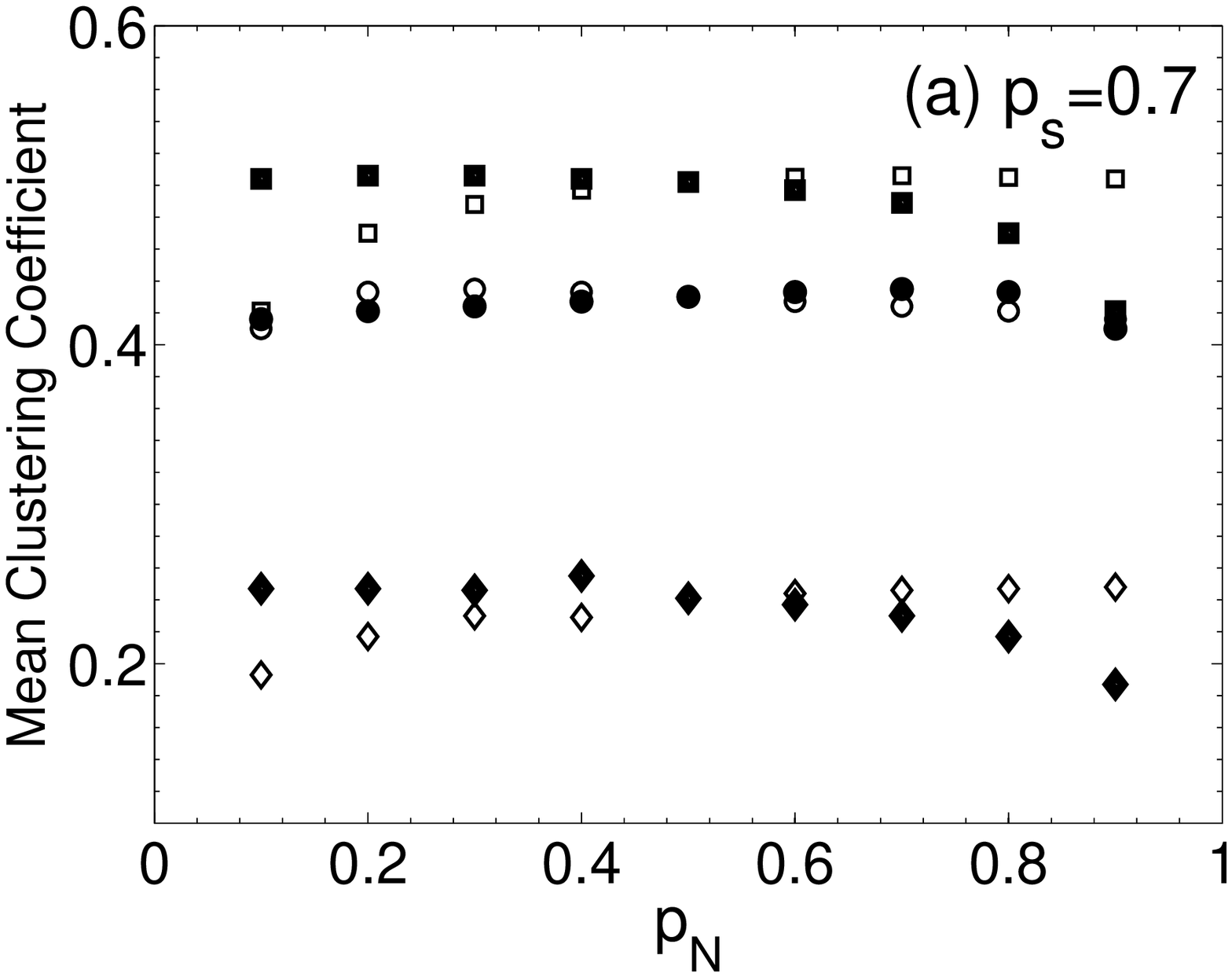}
\hfill
\noindent\includegraphics[width=15pc]{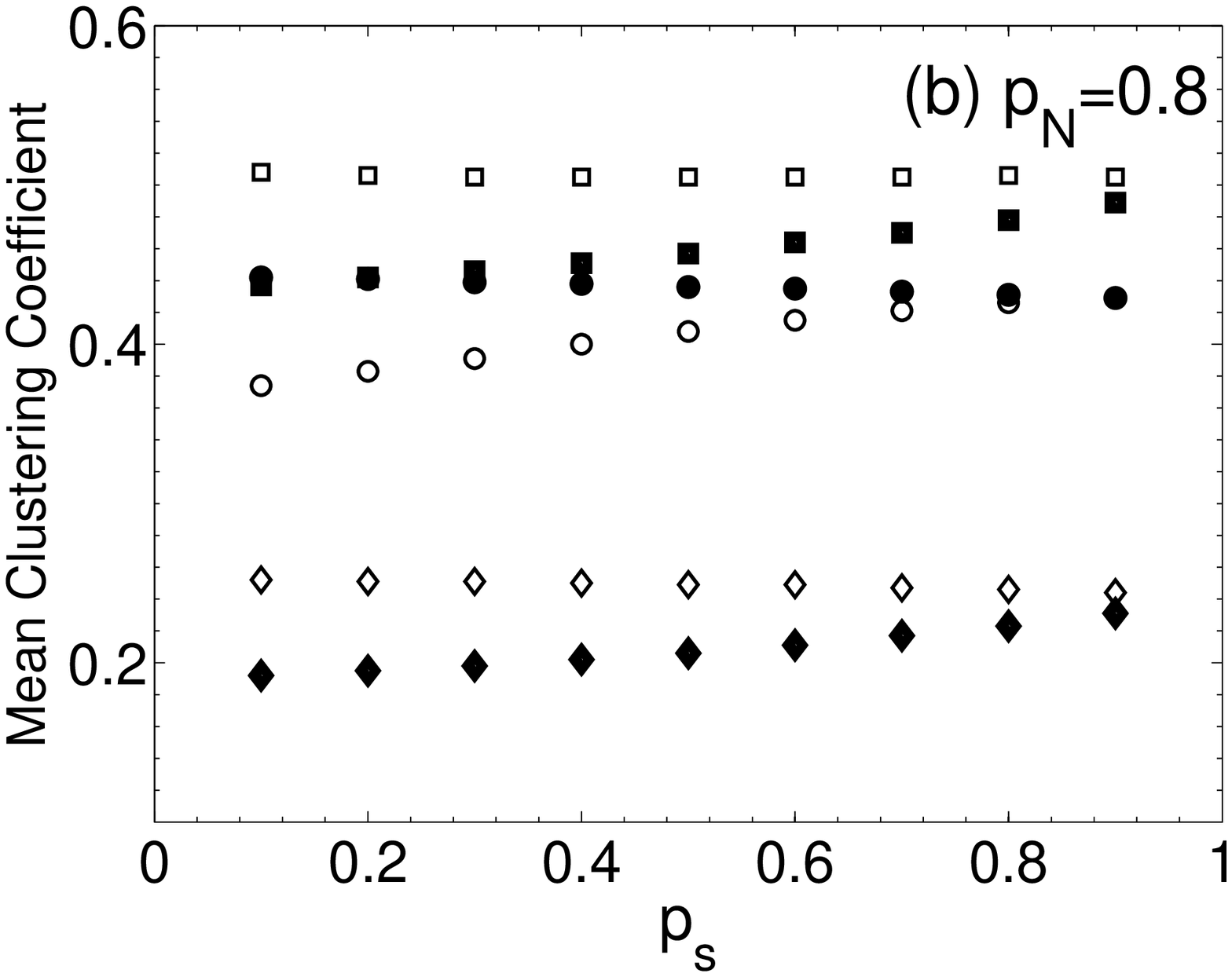}
\vfill
\noindent\includegraphics[width=15pc]{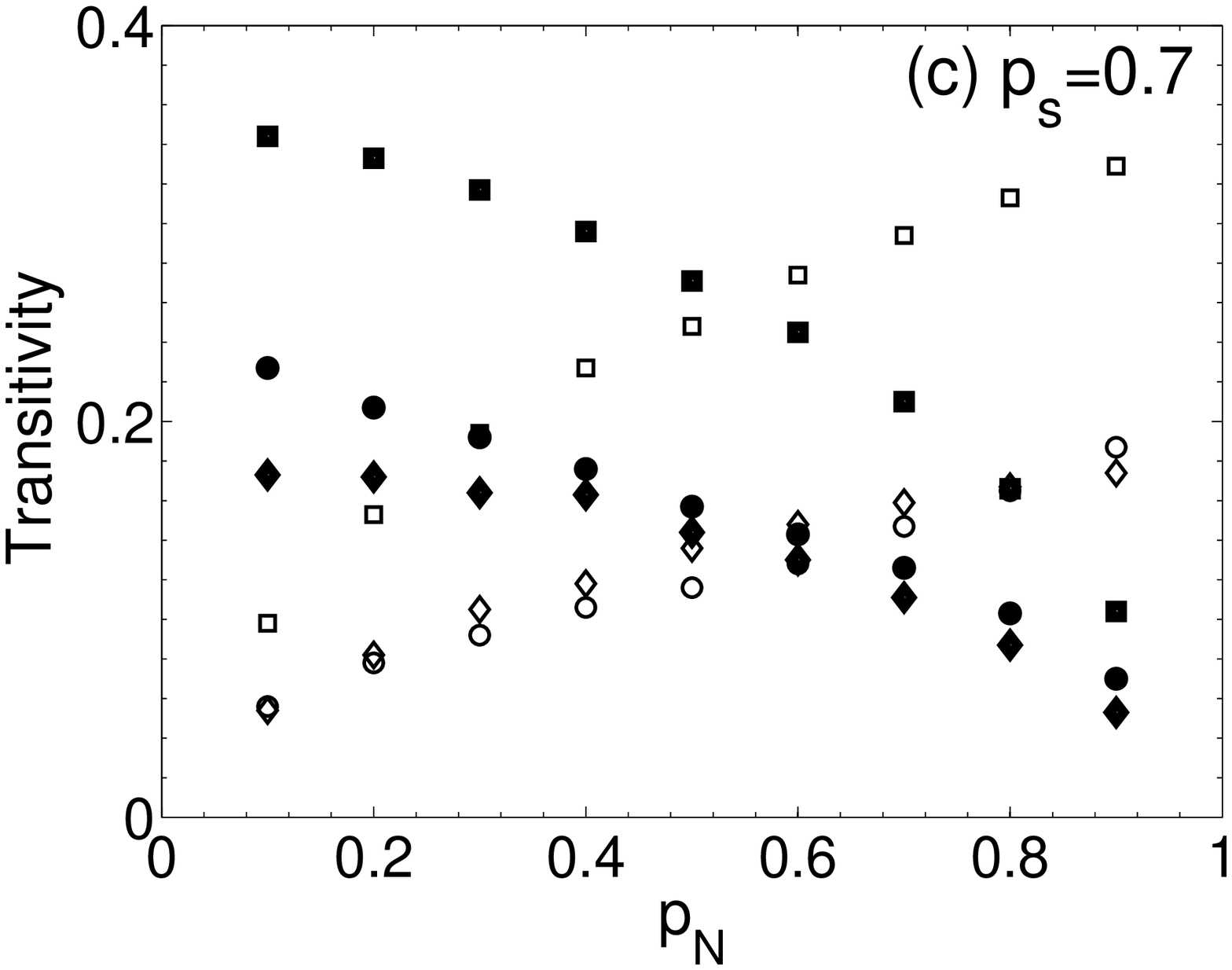}
\hfill
\noindent\includegraphics[width=15pc]{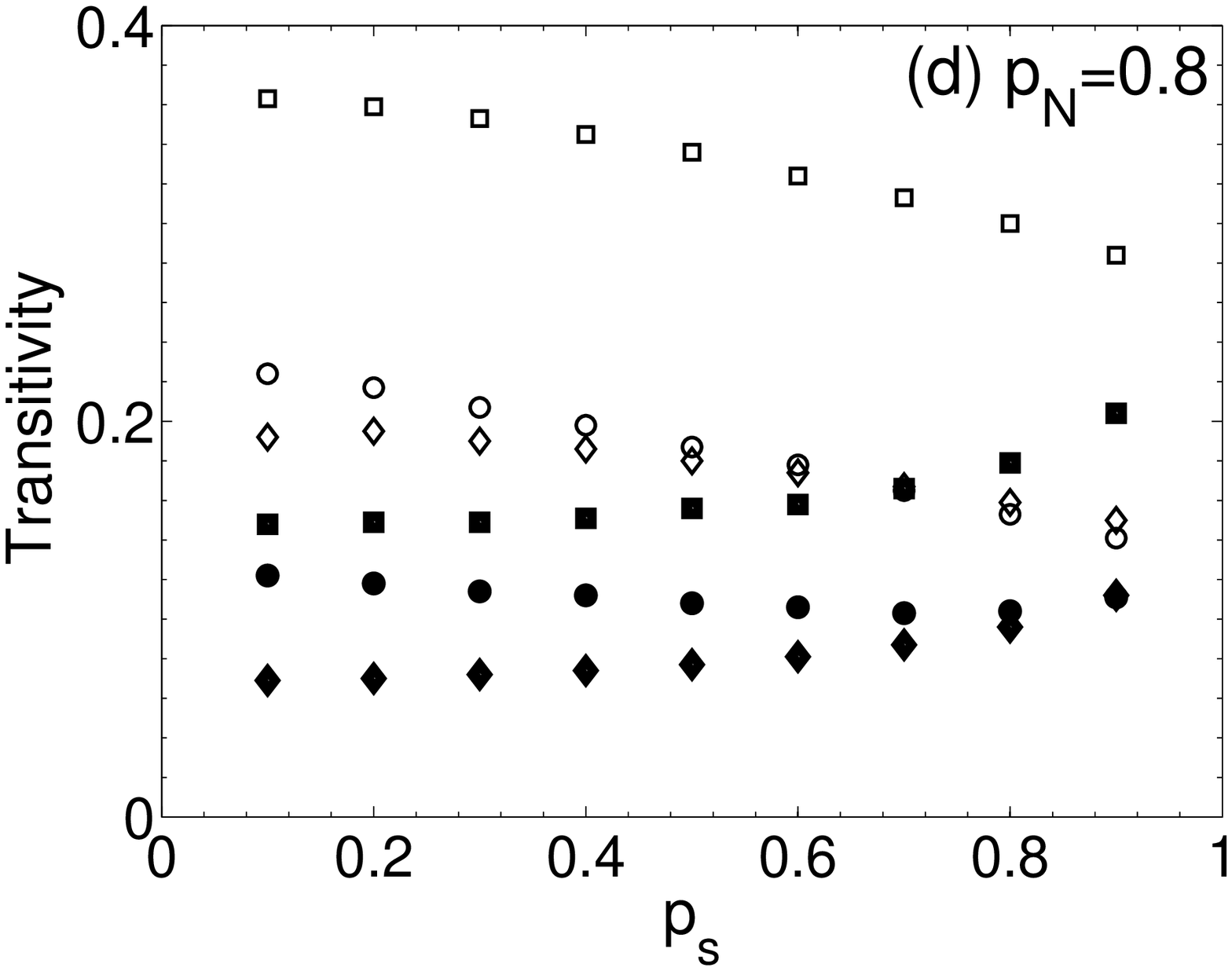}
\caption{Values of the mean clustering coefficient 
$\bar{C}$ (Eq.~(\ref{bc})
(a) as a function of the probability to create an $N$ node $p_N$ for a 
fixed value of the probability to create a link between same type of nodes
$p_s=0.7$ and (b) as a function of $p_s$ for a fixed value of $p_N=0.8$ 
for $N$ nodes (empty symbols) and $V$ nodes (full symbols). 
(c,d) Same as (a,b) but for the values of transitivity $C$ (Eq.~(\ref{T}).
Circles ($N$$\circ$/$V$$\bullet$) represent Case I. 
Squares ($N$$\square$/$V$$\blacksquare$) represent Case II and 
diamonds ($N$$\lozenge$/$V$$\blacklozenge$) mark
Case III. Cases I, II, and III are as defined in Fig. 3.}
\label{fig5}
\end{figure}

The global clustering properties of the network are assessed by the mean
clustering coefficient $\bar{C}$ (Eq.~(\ref{bc})) which is averaged over vertex 
degree, and the transitivity $C$ (Eq.~(\ref{T})).
We study how $\bar{C}$ and $C$ change as a function of $p_N$ for a fixed value 
of $p_s$ (plotted in Fig.~\ref{fig5}a,c) and as a function of $p_s$ for a fixed 
value of $p_N$ (plotted in Fig.~\ref{fig5}b,d). The mean clustering coefficient
$\bar{C}$ for Case I (circles in Fig.~\ref{fig5}a) has values in the range 
between 0.41 and 0.43 as a function of $p_N$ for both $N$ and $V$ nodes
for fixed value of $p_s=0.7$. In both Case II (squares) and Case III (diamonds), 
where the number of secondary contacts is drawn from a uniform distribution, 
the $\bar{C}(p_N)$ for $N$ and $V$ nodes is symmetrical with respect 
to its value at $p_N=0.5$. Values of $\bar{C}$ for $N$ and $V$ nodes as 
a function of $p_s$ (Fig.~\ref{fig5}b) demonstrate a 
tendency to converge for $p_s$ approaching one. 

The transitivity $C$ (Fig.~\ref{fig5}c)
as a function of $p_N$ shows a symmetrical pattern for $N$ and $V$ nodes
with respect to its value at $p_N=0.5$ similar to $\bar{C}$ behavior but with
a wider difference between the results for $N$ and $V$ nodes and different
values. As the probability to establish a link between same type of nodes $p_s$ 
increases the values of transitivity $C$ (Fig.~\ref{fig5}d) for $N$ and $V$
nodes converge. Higher values of both $\bar{C}$ and $C$ among the three
cases are obtained for Case II when there is an option to create either one
or two initial contacts and the number of secondary contacts vary as well, e.g.
in $U[0,3]$. 

\begin{figure}
\noindent\includegraphics[width=15pc]{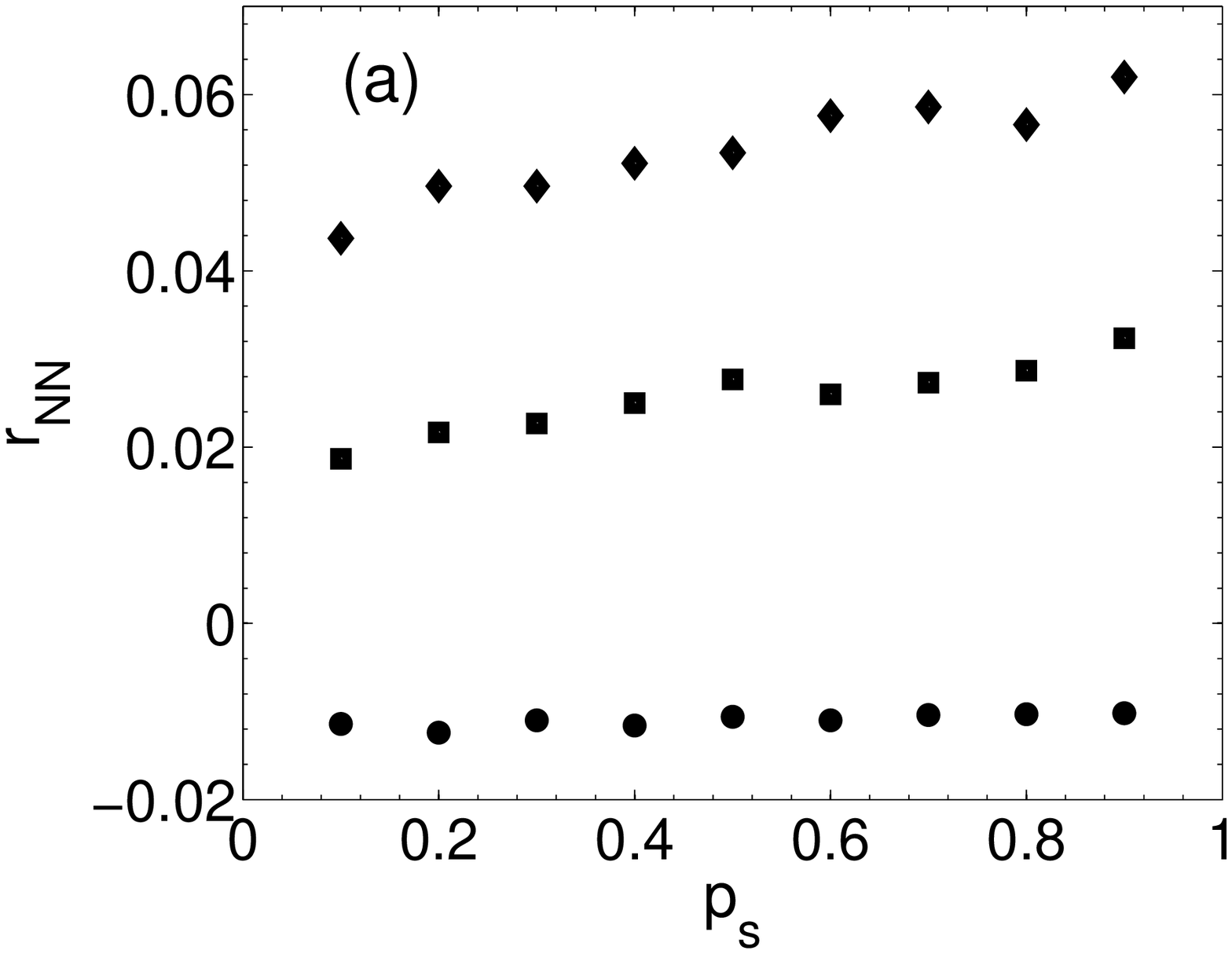}
\hfill
\noindent\includegraphics[width=15pc]{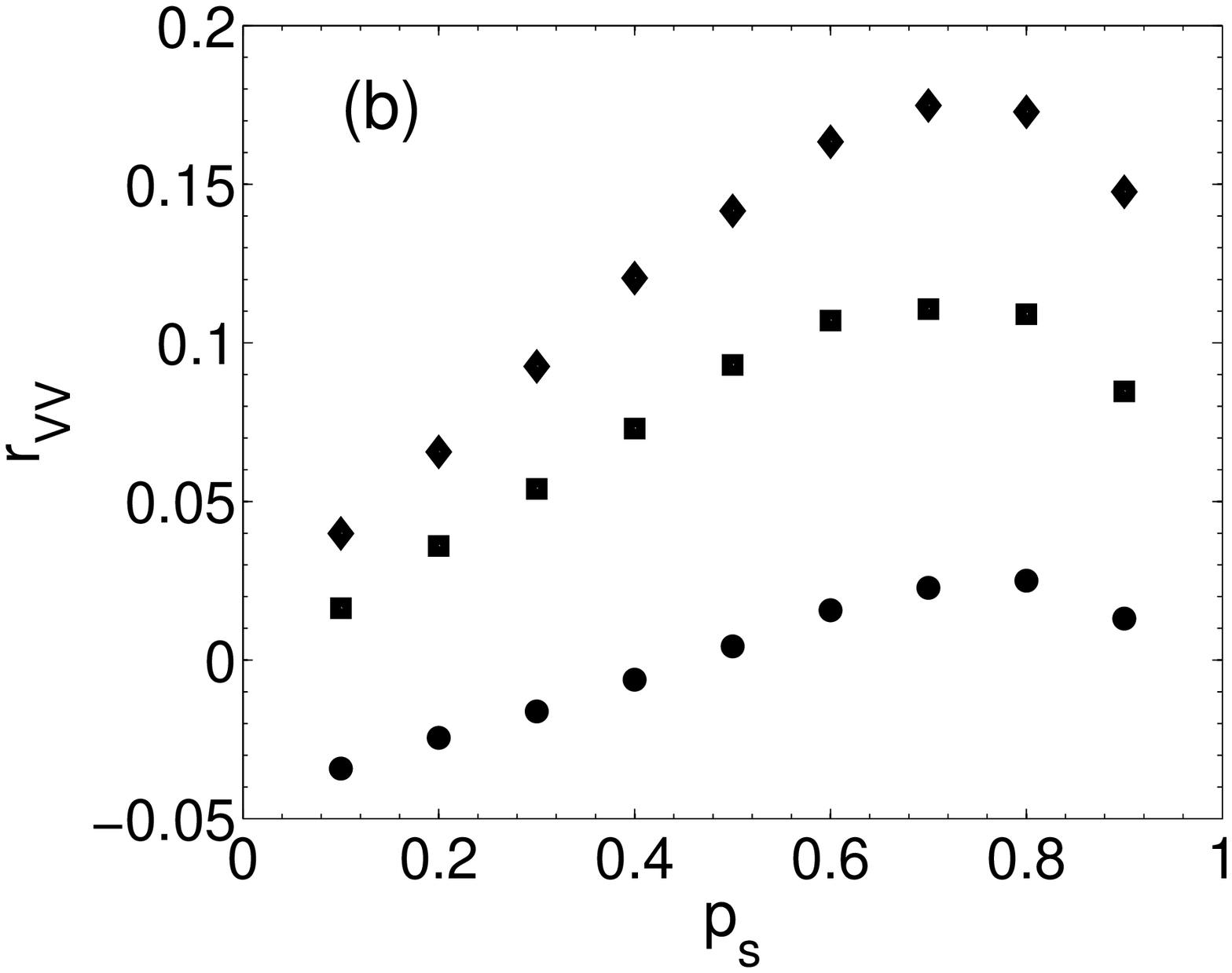}
\vfill
\noindent\includegraphics[width=15pc]{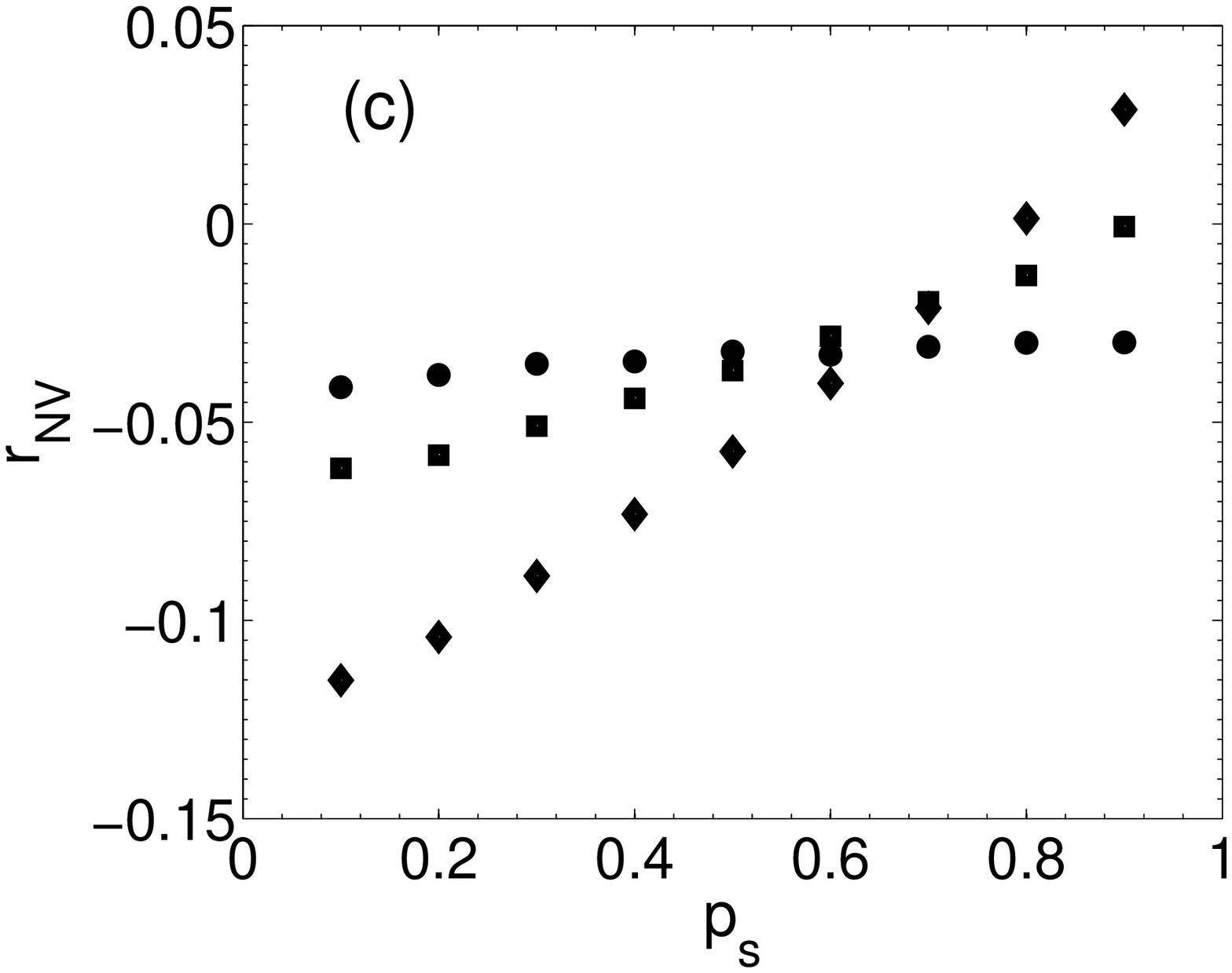}
\caption{Values of assortativity coefficient $r$ as a function 
of the probability to create a link between same type of nodes $p_s$
for a fixed value of $p_N=0.8$ of (a) $NN$ network; (b) $VV$ network;
(c) $NV$ network. Circles ($\bullet$) represent Case I. Squares 
($\blacksquare$) represent Case II and diamonds ($\blacklozenge$) 
mark Case III. Cases I, II, and III are as defined in Fig. 3.}
\label{fig6}
\end{figure}

\subsection{Assortativity}

The assortative properties of the network describe the degree correlation 
of the nodes at the ends of an edge and are quantified by the Pearson
correlation coefficient $r$. The network is said to be assortative when high 
degree nodes tend to connect to other high-degree nodes and $r\in(0, 1]$ 
\cite{newman2002}. 
The network is characterized by disassortative mixing when high degree nodes
tend to connect to low degree nodes and $r\in[-1, 0)$.
We calculate $r$ of each of $NN$, $VV$, and $NV$ networks obtained for a 
fixed value of $p_N=0.8$ and plot the $r$-dependence
as a function of $p_s$ in Fig.~\ref{fig6}a-c for the three cases considered.
$NN$ and $NV$ networks of Case I ($\bullet$) show 
slight disassortative mixing owing to the two initial contacts and one secondary 
contact. Increasing the probability $p_s$ to create an edge between the same 
type of nodes for Case I ($\bullet$ in Fig.~\ref{fig6}b) results in 
changing from disassortative for $p_s<0.5$ to slightly assortative for 
$p_s>0.5$.
Varying the number of initial and secondary contacts in Case II ($\blacksquare$)
and III ($\blacklozenge$) produces
assortatively mixed $NN$ and $VV$ networks (Fig.~\ref{fig6}a,b). 
The assortativity coefficient $r_{VV}$ of the $VV$ networks ($\blacksquare$ 
and $\blacklozenge$ in Fig.~\ref{fig6}b) increases linearly
with increasing the probability to create a link between the same type of 
nodes reaching a plateau around $p_s\sim 0.7$ and then decreasing for $p_s=0.9$
probably because all available nodes are already connected.
The values of $r_{VV}$ are larger versus $r_{NN}$ because of the smaller number 
($p_V=1-p_N=0.2$) of $V$ nodes available for contacts. 
A recent visualization of the connections in a terrorist network such as 
the Global Salafi Jihad depicts them to form an assortative network  
\cite{chen2009}.
For all three cases the $NV$ networks (Fig.~\ref{fig6}c)
show disassortative mixing of their degree only except for very high 
values of $p_s=0.8$ and $0.9$ for Case III ($\blacklozenge$). 
 A recent empirical study of an online social system reports that 
relationships driven by aggression lead to markedly different systemic 
characteristics than relations of a non-aggressive nature \cite{PNAS2010}. 
Assortativity is a characteristic of global properties of the system. 
In agreement with the empirical findings the assortativity of $NV$ network 
(Fig.~\ref{fig6}c) produced by our model which represents relationships driven 
by aggression is clearly different from the assortativity of $NN$ and $VV$ 
networks (Fig.~\ref{fig6}a,b) which are driven by non-aggressive relationships.

\section{Conclusions}

We introduced a model intended to characterize the interactions between two 
distinct populations, which form links more easily within their group than between
groups. We aim to describe the interactions of potentially violent terrorist 
groups within the context of a largely non-violent population, although the same model 
could, in principle, be applied to other non-mainstream social groups. The model is
kept simple enough so that analytical solutions could be derived and compared 
with empirical parameterizations and numerical simulation results. 
 
The model produces networks with relatively high mean clustering coefficient
$\bar{C}$ and transitivity $C$. Their values vary with the balance between
the initial and secondary contacts. This is expected because
of the interplay between the random and preferential attachments of the
initial and secondary connections, respectively.
The assortativity pattern of modeled networks show that the potentially 
violent $VV$ network qualitatively resembles the connectivity pattern in
terrorist networks reported in \cite{chen2009}.
 The assortativity behavior of $NV$ network which is driven by aggression 
is clearly different than the assortativity pattern of $NN$ and $VV$ networks 
which are non-aggressive relationships; a finding which is in agreement 
with the results of recent empirical study of an online social system 
\cite{PNAS2010}.

\end{document}